\documentclass[twocolumn,10pt, superscriptaddress, amsmath,amssymb, aps, prl, floatfix]{revtex4-2}

\usepackage{natbib}
\usepackage[normalem]{ulem}
\usepackage{comment}
\usepackage{graphicx}% Include figure files
\usepackage{dcolumn}% Align table columns on decimal point
\usepackage{bm}% bold math
\usepackage{physics}
\usepackage{mathtools}
\usepackage{tensor}
\usepackage{xcolor}
\usepackage{hyperref}% add hypertext capabilities
\usepackage{graphicx} % Required for inserting images
\usepackage{amsmath}
\usepackage{enumitem}
\usepackage{color}

\definecolor{mgreen}{rgb}{0.1,0.7,0.1}

\DeclareMathOperator*{\argmax}{argmax}

\begin{document}

\title{A Maximum Entropy Conjecture for Black Hole Mergers}
\author{Monica Rincon-Ramirez}
\email[]{mramirez@alumni.psu.edu}
\affiliation{Institute for Gravitation and the Cosmos, The Pennsylvania State University, University Park, PA 16802, USA}
\affiliation{Department of Physics, The Pennsylvania State University, University Park, PA 16802, USA}

\author{Nathan K. Johnson-McDaniel}
\email[]{nkjm.physics@gmail.com}
\affiliation{Department of Physics and Astronomy, The University of Mississippi, University, Mississippi 38677, USA}

\author{Eugenio Bianchi}
\affiliation{Institute for Gravitation and the Cosmos, The Pennsylvania State University, University Park, PA 16802, USA}
\affiliation{Department of Physics, The Pennsylvania State University, University Park, PA 16802, USA}

\author{\\Ish Gupta}
\affiliation{Department of Physics, University of California, Berkeley, CA 94720, USA}
\affiliation{Department of Physics and Astronomy, Northwestern University, 2145 Sheridan Road, Evanston, IL 60208, USA}
\affiliation{Center for Interdisciplinary Exploration and Research in Astrophysics (CIERA), Northwestern University, 1800 Sherman Ave, Evanston, IL 60201, USA}
\affiliation{Institute for Gravitation and the Cosmos, The Pennsylvania State University, University Park, PA 16802, USA}
\affiliation{Department of Physics, The Pennsylvania State University, University Park, PA 16802, USA}

\author{Vaishak Prasad}
\affiliation{Institute for Gravitation and the Cosmos, The Pennsylvania State University, University Park, PA 16802, USA}
\affiliation{Department of Physics, The Pennsylvania State University, University Park, PA 16802, USA}

\author{B. S. Sathyaprakash}
\affiliation{Institute for Gravitation and the Cosmos, The Pennsylvania State University, University Park, PA 16802, USA}
\affiliation{Department of Physics, The Pennsylvania State University, University Park, PA 16802, USA}
% \affiliation{Department of Astronomy and Astrophysics, The Pennsylvania State University, University Park, PA 16802, USA}

\begin{abstract}
The final state of a binary black hole merger is predicted with high precision by numerical relativity, but could there be a simple thermodynamic principle within general relativity that governs the selection of the remnant? Using post-Newtonian relations between the mass $M$ (including the binding energy) and angular momentum $J$ of quasi-circular, nonspinning binaries, we uncover a puzzling result: When the binary’s instantaneous $M$ and $J$ are mapped to those of a hypothetical Kerr black hole, the corresponding entropy exhibits a maximum during the evolution. This maximum occurs at values of $M$ and $J$ strikingly close to those of the final remnant predicted by numerical relativity. Consistent behavior is observed when using the relation between $M$ and $J$ obtained from numerical relativity evolution.  Although this procedure is somewhat ad hoc, the agreement between the masses and spins of the final state obtained from numerical relativity and the results of this maximum entropy procedure is remarkable, with agreement to within a few percent when using either post-Newtonian or numerical relativity results for $M$ and $J$. These findings allow us to propose an \emph{entropy maximization conjecture} for binary black hole mergers, hinting that thermodynamic principles may govern the selection of the final black hole state.
\end{abstract}

\maketitle

\emph{Introduction}—The merger of two black holes is one of the most remarkable predictions of general relativity, now routinely observed through gravitational waves \cite{LIGOScientific:2025slb}.  Numerical relativity (see, e.g.,~\cite{Baumgarte:2010ndz}) provides a robust framework for computing the dynamics of the merger (see, e.g., the extensive catalog of simulations from the SXS collaboration~\cite{Scheel:2025jct}). For two equal-mass, non-spinning black holes that are initially in a quasi-circular orbit, numerical relativity (NR) predicts \cite{Scheel:2008rj} for the remnant black hole a final mass $M_f$ equal to $(95.162 \pm 0.002)\%$ of the sum of the initial Christodoulou masses of the binary, the rest being radiated away as gravitational waves, and a final dimensionless spin~\footnote{We work in units with $c=1,$ $G=1,$ and $k_B=1$, keeping track of Planck's constant $\hbar$ to distinguish quantum from classical phenomena.} $\chi_f = 0.68646 \pm 0.00004.$ While these results are impressively precise, they invite investigations into whether one can obtain them from a deeper principle. 

The prevailing view in the literature (see, e.g.,~\cite{Buonanno:2007sv,LIGOScientific:2016vlm}) is that the remnant black hole spacetime inherits the binding energy and angular momentum of the binary (purely orbital angular momentum in the nonspinning case) at the point of plunge, and then radiates additional energy and angular momentum during ringdown. But this merely shifts the question: Why does the plunge occur when it does? Do thermodynamic properties of black holes \cite{Davies:1977bgr} play a role in the selection of the final state?  There are phenomenological fits for the mass and spin of the remnant in terms of the masses and spins of the progenitor black holes (see, e.g.,~\cite{Hofmann:2016yih,Healy:2016lce,Jimenez-Forteza:2016oae,PhysRevResearch.1.033015}) which work well but lack a deeper physical rationale.

The objective of this work is to explore the connection between thermodynamics and black hole mergers. It is well known that the laws of black hole mechanics admit a profound thermodynamic interpretation \cite{Bardeen:1973gs}. While the notions of temperature and entropy of a black hole involve quantum phenomena, as evidenced by their dependence on the Planck constant $\hbar$, thermodynamic arguments can highlight non-trivial classical properties of general relativity. For instance, the
Bekenstein-Hawking relation between entropy and horizon area \cite{Bekenstein:1973ur}, together with the second law of thermodynamics, immediately implies that in a black hole merger the area of the horizon cannot decrease---a non-trivial prediction of general relativity \cite{Hawking:1971tu,Ashtekar:2025wnu,LIGOScientific:2025rid}. The merger of two black holes is a far-from-equilibrium process and is an open system due to the emission of gravitational waves. Nevertheless, it can be compared to the thermalization of two bodies initially at different temperatures that are brought into thermal contact. For the latter system, the final equilibrium state can be predicted using the principle of maximum entropy \cite{fermi1956thermodynamics,callen1985thermodynamics}: The two bodies reach thermal equilibrium once the final temperature $T_f$ maximizes the entropy at fixed total energy. Although this analogy in classical thermodynamics is not exact, we begin by presenting a puzzling result in post-Newtonian (PN) theory that raises the question of whether a maximum entropy principle governs the final state of a black hole merger.

Throughout this Letter, we use $M_{1,2}$ to denote the Christodoulou and point masses of the component BHs in the context of NR simulations and the PN approximation, respectively, in both cases at the beginning of the inspiral; $M_\mathrm{tot} = M_1 + M_2$ for the total initial mass; and $M, J$ to denote the instantaneous mass (including the binding energy) and angular momentum of the system, respectively, in the PN context, and the Bondi mass and angular momentum, respectively, in the NR context; $M_0, J_0$ denote their initial values.

\emph{PN expansion and Kerr entropy}—The PN expansion is a perturbative framework for describing the inspiral of compact binaries within general relativity. The PN formalism systematically incorporates relativistic corrections in powers of $v$, where $v$ is the characteristic velocity of the system. While highly successful at modeling the conservative dynamics and radiation reaction of binary black holes during inspiral (treated as a binary of point particles), the PN expansion lacks an
intrinsic mechanism for determining its domain of validity. In particular, it does not provide a definitive prescription for when the inspiral ends and the merger begins. As such, the termination criterion is an open question that must be determined by some other physics \cite{Blanchet:2024PN}, e.g., typically by empirical thresholds such as the last stable orbit \cite{Blanchet:2025agj}, or the minimum energy orbit \cite{Cabero:2016ayq}, or by comparison with NR simulations. In this work, we stop the PN evolution at the point of maximum entropy described below.

Within the PN framework, physical quantities characterizing a black hole binary can be expressed as a series expansion in a small parameter~\footnote{It is important to note that the post-Newtonian expansion is an asymptotic series that is not guaranteed to monotonically approach its limiting value \cite{Blanchet:2024PN, Damour:1997ub}.}, typically chosen as the dimensionless parameter $x \equiv v^2 = (M_\mathrm{tot} \Omega)^{2/3},$  where $\Omega$ is the orbital frequency of the binary.  For instance, the orbital angular momentum $J(x)$, the binding energy $E(x)$, and the mass $M(x)=M_\mathrm{tot}+E(x)$ of the binary system, can all be expressed in terms of the orbital frequency $\Omega$ as an expansion in the dimensionless variable $x$, with $0< x\ll1$. Because of the emission of gravitational radiation, $M$ and $J$ both decrease monotonically in time, corresponding to an increasing orbital frequency $\Omega$; the companion masses $M_1$, $M_2$ remain unchanged in this approximation. Therefore, we can express how the mass of the binary depends on its orbital angular momentum by deparametrizing the curve $(M(x),J(x))$ and using $J$ as a time parameter. Working self-consistently at the fourth order in the PN expansion (4PN), where $n$PN corresponds to including terms through $O(x^n)$ beyond Newtonian gravity, one finds the expression for $M(j)$ \cite{4PN_Damour} 

\begin{align}
 M& =  M_\mathrm{tot} -\frac{\nu M_\mathrm{tot}}{2j^2}\left\{1+\left ( \frac{9+\nu}{4} \right ) \frac{1}{j^2}+\frac{81-7\nu+\nu^2}{8}\frac{1}{j^4}\right. \nonumber \\
    & +\left[\frac{3861}{64}+\left(\frac{41\pi^2}{32}-\frac{8833}{192}\right)\nu-\frac{5\nu^2}{32}+\frac{5\nu^3}{64}\right]\frac{1}{j^6}\nonumber\\
    & +\left[\left(\frac{6581\pi^2}{512}-\frac{989911}{1920}\right. -\frac{128}{5} \left ( \gamma_E  + \ln \frac{4}{j} \right ) \right)\nu\nonumber \\ & + \left(\frac{8875}{384}-\frac{41\pi^2}{64}\right)\nu^2 \left.\left.-\frac{3\nu^3}{64}+\frac{7\nu^4}{128}+\frac{53703}{128}\right]\frac{1}{j^8}\right\},  
    \label{eq:Mj}
\end{align}
where $\nu=M_1 M_2/M_\mathrm{tot}^2$ is the symmetric mass ratio, and we have introduced a dimensionless parameter $j,$ as customary, for the orbital angular momentum $J=M_\mathrm{tot}^2\,\nu j$. This formula bypasses the PN parameter $x$ and allows us to study how the mass evolves as a function of the angular momentum parameter $j$, which serves as an alternative time variable. 

To state the main observation of this paper, we need a second ingredient: the entropy of a rotating black hole. A Kerr spacetime \cite{Kerr:1963ud} with mass $M$ and angular momentum $J$ describes the equilibrium configuration of a rotating black hole in vacuum general relativity. The black hole horizon has an area $A(M,J)$ given by the formula
\begin{equation}\label{eq:area}
A(M,J)=8\pi\Big(1+\sqrt{1-\tfrac{J^2}{M^4}}\,\Big)\,M^2\,.
\end{equation} 
The Bekenstein-Hawking formula states that the entropy of the black hole is proportional to the horizon area and is given by $S=A(M,J)/4\hbar$ \cite{Bekenstein:1973ur}.
For a binary system with mass $M$ and angular momentum $J$, we will consider the entropy $S_\mathrm{Kerr}$ of an analogue Kerr spacetime with the same mass $M$ and spin $J$.

\begin{figure}[t!]
    \centering    \includegraphics[width=\linewidth]{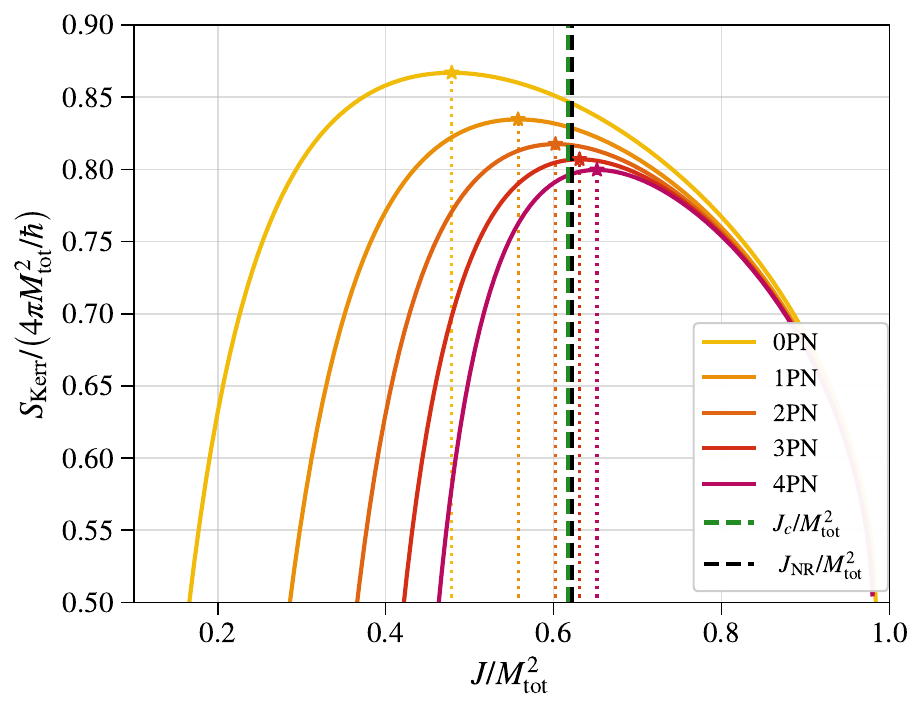}
    \caption{
        Entropy $S_\mathrm{Kerr}(J)$ of the remnant, normalized to $4\pi M_\mathrm{tot}^2/\hbar,$ as a function of the normalized angular momentum $J/M_\mathrm{tot}^2 = j\nu,$ for an equal-mass binary. At each instant of PN evolution, one may consider a hypothetical remnant inheriting the angular momentum and energy of the binary. While the PN formalism does not, by itself, determine the ultimate outcome of the evolution, the \emph{maximum-entropy conjecture} provides a natural criterion for its termination: For a given PN order, the conjecture dictates that we should end the sequence of circular orbits at the value of angular momentum that maximizes the entropy of the remnant, $J_* = \argmax_J S_\mathrm{Kerr}(J)$, marked with a star for each PN curve.  Reference values of $J/M_\mathrm{tot}^2$ are displayed as vertical lines, corresponding to (black) the value of final angular momentum predicted by numerical relativity $J_\mathrm{NR}=\chi_{\mathrm{NR}}M_{\mathrm{NR}}^2$ \cite{Scheel:2008rj}, and (green) the value of angular momentum for which the specific heat $C_J$ diverges, $J_\mathrm{NR}=\chi_{\mathrm{c}}\,M_{\mathrm{NR}}^2$, for a black hole of mass $M_\mathrm{NR}=0.9516\, M_\mathrm{tot}$, where  $\chi_c = \sqrt{2\sqrt3-3}\approx 0.681$ \cite{Davies:1977bgr}. 
    }
    \label{fig:entropy vs spin}
\end{figure}

As the binary system evolves, it traces out a one-parameter family of quasi-equilibrium states, each state parameterized by the value of $J$ and corresponding mass $M$. When does the sequence of PN orbits end to produce a black hole remnant? The PN framework is agnostic to the precise merger condition. Consequently, we may consider each $j$ along a sequence of values as a potential candidate for the endpoint of the inspiral.  Figure \ref{fig:entropy vs spin} shows the entropy 
\begin{equation}
S_\mathrm{Kerr}(j)\equiv \frac{A(M(j),J(j))}{4\hbar} \label{eq:Sofj}
\end{equation}
as a function of $J/M_\mathrm{tot}^2 = j\nu$ at different PN orders for hypothetical remnants with final spin parameter $j$. Using the PN expansion~(\ref{eq:Mj}) for $M(j)$, we see that the Kerr horizon area (\ref{eq:area}) can be expressed as
\begin{equation}
\frac{A(M(j),J(j))}{8\pi M_\mathrm{tot}^2}%\approx
= \Big[1+\sqrt{1-(\nu j)^2}\Big]\left[1 + O\left(\frac{1}{j^2}\right)\right]
%+\textrm{1PN}+\cdots
\,,
\end{equation}
and becomes a real quantity for $j\leq \nu^{-1}$ ($\Rightarrow J/M_\mathrm{tot}^2 \leq 1$). 
We interpret the cases where the area is imaginary (because the dimensionless spin is greater than 1) as indicating that it is too early in the evolution to possibly form a final Kerr black hole.

As the binary loses energy and angular momentum, the spin parameter $j$ decreases and the entropy $S_\mathrm{Kerr}(j)$ initially increases, reaching a maximum for a specific value of $j_*$, beyond which further evolution decreases the entropy. The key observation is that the mass $M(j_*)$ and the spin $J(j_*)$ at the point of maximum entropy determine a final spin parameter $\chi_*\equiv J(j_*)/M(j_*)^2$ which is remarkably close to the one determined in numerical relativity simulations, $\chi_f = 0.68646 \pm 0.00004$ \cite{Scheel:2008rj}.  Fig.~\ref{fig:entropy vs spin} and Table~\ref{tab:chif-epsilonf} show the values of $\chi_*$ obtained at different PN orders for an equal mass binary black hole (i.e., $\nu=1/4$), compared to the value $\chi_f$.

Thus, somewhat puzzlingly, the PN results already appear to be consistent with the remnant approaching a state that maximizes entropy, even though the PN approximation is formally valid only in the weak-field, slow-motion regime. This apparent agreement with a property of the final state resulting from a strongly nonlinear evolution suggests either an unexpected imprint of the merger outcome in the PN dynamics, or a coincidence that warrants deeper theoretical understanding.

\emph{Maximum Entropy Conjecture for Black Hole Mergers}
In order to formulate a \emph{maximum entropy conjecture} for black hole mergers, we need to clarify the definition of two ingredients: (i) how the entropy of a remnant candidate is defined, and (ii) what the balance laws for the energy and the angular momentum are.

\begin{figure*}[htb!]
    \centering    \includegraphics[width=0.49\linewidth]{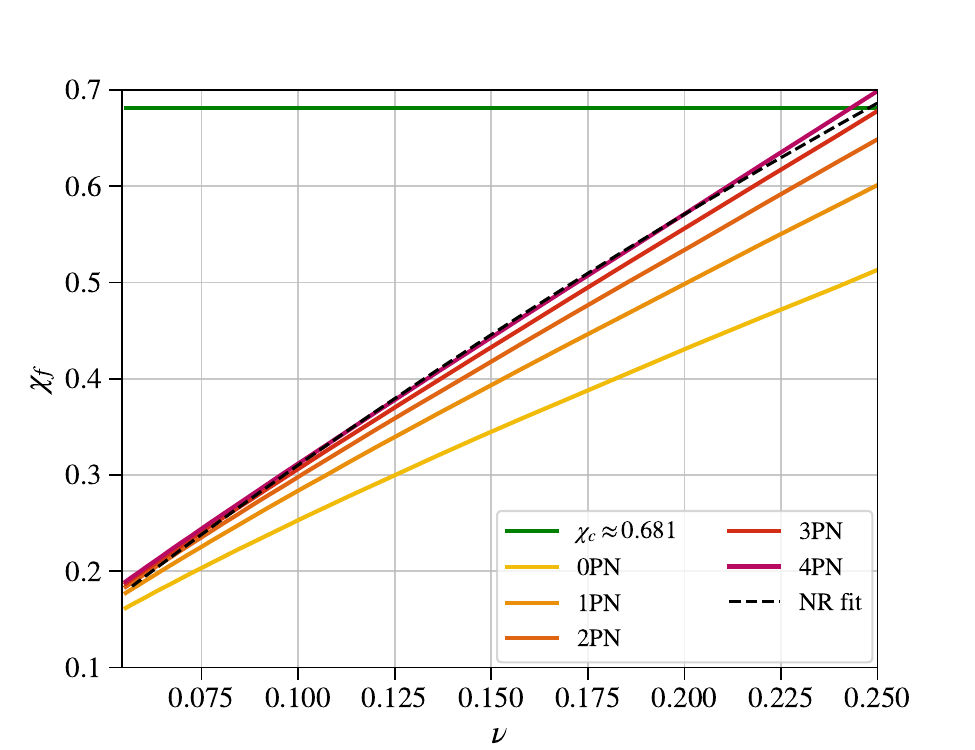}    \includegraphics[width=0.458\linewidth]{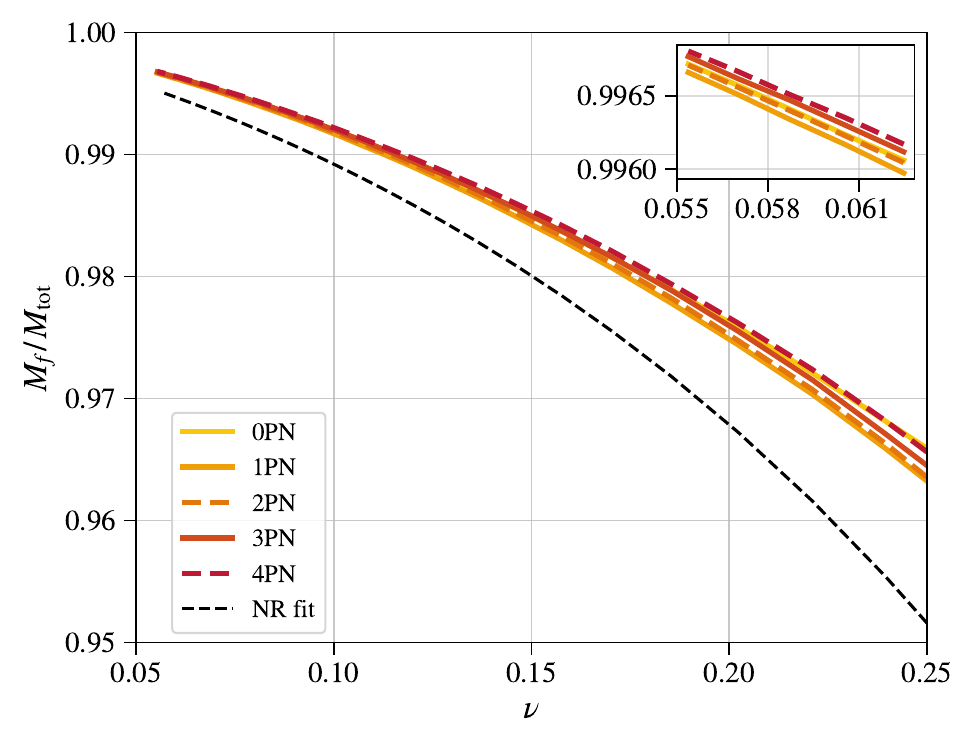}
    \caption{The spin $\chi_f$ (left) and mass $M_f$ (right) of the remnant black hole obtained from the integral form of the maximum entropy conjecture applied using the PN approximation at different PN orders. Also shown as a dashed line are the predictions of NR simulations. The $4$PN remnant spin is pretty close to the one predicted by full general relativity.}
    \label{fig:remnant mass and spin}
\end{figure*}

To address point (i), consider the merger of two black holes with masses $M_1$ and $M_2$, both initially non-spinning and in a quasi-circular inspiraling orbit, characterized by an initial orbital angular momentum $J_0$. As the system evolves, gravitational-wave emission leads to a loss of both energy and angular momentum. The merger eventually culminates in the formation of a remnant---a Kerr black hole with final mass $M_f$ and angular momentum $J_f$. General relativity predicts that, in this process, the area of the dynamical horizon is non-decreasing \cite{Ashtekar:2004cn, Ashtekar:2025wnu}, as shown also by numerical simulations \cite{Scheel:2014ina, Gupta:2018znn} (more generally than dynamical horizons). Correspondingly, the entropy of the binary system is initially given by the sum of the entropies of the two black holes and remains approximately constant during the inspiral phase, until the merger, when it rapidly increases and approaches the final equilibrium value given by the entropy of the Kerr black hole with parameters $M_f$ and $J_f$. Once this configuration is approached, small perturbations satisfy the first law of black hole mechanics \cite{Bekenstein:1973ur}, which states:
%\nkjm{We should presumably have a reference for the form given here.} 
\begin{equation}
dS=\frac{1}{T_H}\big(dM-\Omega_H \,dJ\big)\,,
\end{equation}
where $T_H=\hbar \kappa_H/2\pi$ is the Hawking temperature, $\kappa_H$ is the surface gravity, and $\Omega_H$ is the horizon angular velocity of the equilibrium black hole, understood as functions of $M$ and $J$. Specifically,
\begin{equation}
\Omega_H(M,J)=\frac{1}{2M}\frac{J/M^2}{1 + \sqrt{1-J^2/M^4}}\,.
\end{equation}
In particular, we have the relation
\begin{equation}
\frac{dM}{dJ}=\Omega_H\,,
\label{eq:dMdJ-H}
\end{equation}
which applies to  perturbations that do not change the entropy of the remnant Kerr black hole, $dS=0$. Note that, even though $S$ and $T_H$ depend on  $\hbar$, it cancels out, and the expression has a clear classical interpretation.

%-eb- new version of the table
\setlength{\tabcolsep}{9pt} % default is usually 6pt; increase for wider spacing
\renewcommand{\arraystretch}{1.25}  % 25% more vertical space
\begin{table}[b]
    \centering
    \begin{tabular}{c|c|c|c|c|c}
        & 0PN & 1PN & 2PN & 3PN & 4PN \\
        \hline
        $\chi_f$ & 0.513 & 0.601 & 0.649 & 0.678 & 0.699 \\
        $\epsilon_f$ & 3.4\% & 3.6\% & 3.6\% & 3.5\% & 3.4\% \\
        $\tfrac{M_f}{M_\mathrm{tot}}$ & 0.966 & 0.963 & 0.964 & 0.965 & 0.966 \\
         % $\left.\frac{J}{M^2}\right|_\mathrm{\small MECO}$ & -- &0.674 &0.811 &0.817 &0.828 
        $\chi^{}_{{\scriptscriptstyle \mathrm{MECO}}}$ & -- &0.674 &0.811 &0.817 &0.828 
    \end{tabular}
    \caption{Remnant dimensionless spin $\chi_f \equiv J_f/M_f^2$, fractional radiated energy $\epsilon_f \equiv 1- M_f/M_\mathrm{tot}$, and remnant mass $M_f$ as a fraction of the initial total mass $M_\mathrm{tot}$, computed at various PN orders with Eq.~\eqref{eq:Mj} and the integral version of the maximum entropy conjecture, Eq.~\eqref{eq:argmaxJ}. At higher PN orders, $\chi_f$ determined by the conjecture is close to the general relativistic value of $\chi_f = 0.68646 \pm 0.00004$ from numerical relativity simulations~\cite{Scheel:2008rj};  $\epsilon_f$ (and thus $M_f/M_\mathrm{tot}$) is roughly the same at all PN orders but smaller than the general relativistic value of $(4.838 \pm 0.002)\%$~\cite{Scheel:2008rj}; this is possibly due to strong sensitivity of the radiated energy to the final value of the angular momentum; even small deviations in $j_f$ cause large shifts in $\epsilon_f.$ Finally, $\chi^{}_{\scriptscriptstyle\mathrm{MECO}} \equiv \left.(J/M^2)\right|_{\scriptscriptstyle\mathrm{MECO}}$ denotes the dimensionless spin of the remnant, if it were to form at the minimum-energy circular orbit (MECO), defined by $dE/dx = 0$ \cite{Cabero:2016ayq}, and significantly differs from the NR prediction at higher PN orders.}
    \label{tab:chif-epsilonf}
\end{table}

% %- original table
% \setlength{\tabcolsep}{9pt} % default is usually 6pt; increase for wider spacing
% \begin{table}[b]
%     \centering
%     \begin{tabular}{c|c|c|c|c|c}
%         & 0PN & 1PN & 2PN & 3PN & 4PN \\[1pt]
%         \hline
%         $M_f/M_{tot}$ & 0.966 & 0.963 & 0.964 & 0.965 & 0.966 \\
%         $\chi_f$ & 0.513 & 0.601 & 0.649 & 0.678 & 0.699 \\
%         $\epsilon_f$ & 3.4\% & 3.6\% & 3.6\% & 3.5\% & 3.4\% \\
%     \end{tabular} 
%     \caption{The remnant mass $M_f$ as a fraction of the initial total mass $M_{tot},$ remnant dimensionless spin $\chi_f \equiv J_f/M_f^2$, and fractional radiated energy $\epsilon_f \equiv 1- M_f/M_{tot}$ computed at various PN orders with Eq.~\eqref{eq:Mj} and the integral version of the maximum entropy conjecture, Eq.~\eqref{eq:argmaxJ}. In contrast, Fig.~\ref{fig:remnant mass and spin} uses the differential version of the conjecture. At higher PN orders, $\chi_f$ determined by the conjecture is close to the general relativistic value of $\chi_f = 0.68646 \pm 0.00004$ from numerical relativity simulations~\cite{Scheel:2008rj};  $\epsilon_f$ (and thus $M_f/M_{tot}$) is roughly the same at all PN orders but smaller than the general relativistic value of $(4.838 \pm 0.002)\%$~\cite{Scheel:2008rj}; this is possibly due to strong sensitivity of the radiated energy to the final value of the angular momentum; even small deviations in $j_f$ cause large shifts in $\epsilon_f.$} 
%     \label{tab:chif-epsilonf}
% \end{table}

To address point (ii), we consider balance laws for the flux of energy and angular momentum at future null infinity $\mathcal{I}^+$. As the system is open, and gravitational waves carry energy and angular momentum to infinity, we need to take into account balance laws, instead of conservation laws (such as fixed total energy for the two isolated bodies in thermodynamics). In general relativity, the balance laws relate the Bondi fluxes of energy $\mathcal{F}_E(u)$ and of angular momentum $\mathcal{F}_J(u)$ radiated away to null infinity, to the Bondi mass $M$ and angular momentum $J$ of the system  at the retarded time $u$~\cite{M_dler_2016}:
\begin{equation}
\mathcal{F}_E(u)=-\frac{d M(u)}{du}\,,\quad\; \mathcal{F}_J(u)=-\frac{d J(u)}{du}\,.
\end{equation}
% \nkjm{We probably need a reference for the Bondi quantities.} \bss{These are just  sabalance laws saying flux of energy comes from time variation of energy and flux of momentum from time variation of momentum, we really don't need any reference here; they are very generic laws.}
As the angular momentum $J(u)$ of the system decreases monotonically, we can use $J$ as a time parameter in place of $u$, and express the mass of the system as $M(J)$.

Integrating these equations from the initial conditions at $u = u_0$, we obtain the integral form for the final mass $M_f$ and spin $J_f:$ 
\begin{equation}
M_f=M_0-\int_{u_0}^\infty\!\!\!\mathcal{F}_E(u)du\,,\;\;\;
J_f=J_{0} -\int_{u_0}^{\infty}\!\!\!\mathcal{F}_J(u)du\,,
\label{eq:Mf and Jf integrals}
\end{equation}
where $u_0$ is the earliest retarded time and $M_{0} = M(u_0)$ and $J_{0}=J(u_0)$.

For a binary black hole system in a quasi-circular orbit, there is an additional approximately conserved quantity. Consider a spacetime with a global helical Killing vector \cite{Friedman:2001pf} $k^\mu=t^\mu+\Omega_{\mathrm{orb}}\,\phi^\mu$, where $\Omega_{\mathrm{orb}}>0$ is a constant, $t^\mu$ is timelike, and $\phi^\mu$ is spacelike with circular orbits of length $2\pi$. Neither $t^\mu$ nor $\phi^\mu$ is a Killing vector, but the combination $k^\mu$ is an approximate Killing vector of the binary system with $\Omega_{\mathrm{orb}}$ representing the orbital angular velocity of the binary. There is an approximate conserved quantity $Q$ associated with the helical Killing vector $k^\mu$. The requirement that its variation vanishes $0=dQ=dM-\Omega_{\mathrm{orb}}\, dJ$ implies the additional balance law~\cite{LeTiec:2017ebm}
\begin{equation}
\frac{dM}{dJ}=\frac{\mathcal{F}_E(u)}{\mathcal{F}_J(u)}=\Omega_{\mathrm{orb}}(u)\,.
\label{dMdJ-orb}
\end{equation}
This expression is exact, order by order in the PN expansion, where it takes the form $dM/d\Omega=\Omega\, dJ/d\Omega$ with the orbital angular velocity $\Omega=x^{3/2}/M_\mathrm{tot}$. Moreover, after the merger and once the remnant black hole reaches equilibrium, both the vectors $t^\mu$ and $\phi^\mu$ become Killing vectors, and the helical Killing vector $k^\mu$ defines the horizon angular velocity $\Omega_H$.

With these clarifications, we are now ready to state the maximum entropy conjecture for the remnant in a  black hole merger: Given the function $M(J)$ determined by the balance laws, the final mass and spin of the black hole remnant are $M_f=M(J_*)$ and $J_f=J_*$ with the value $J_*$ where the area of a Kerr black hole with mass $M(J)$ and spin $J$ reaches a maximum:
\begin{equation}\label{eq:argmaxJ}
J_*=\argmax_J  A(M(J),J)\,.
\end{equation}
This integral formulation admits an equivalent differential formulation. Since the stationarity condition $dA(M(J),J)/dJ=0$ implies  \eqref{eq:dMdJ-H}, using \eqref{dMdJ-orb}, we find that the final spin $J_f$ satisfies the condition
\begin{equation}\label{eq:diff_form}
\Omega_\mathrm{orb}(J_f)=\Omega_H(M(J_f),J_f)\,.
\end{equation}
When we use full general relativity, the differential and the integral versions in \eqref{eq:dMdJ-H} and~\eqref{eq:argmaxJ} are exactly equivalent, while \eqref{eq:diff_form} requires also assuming approximate helical symmetry. 
If we use, instead, the PN expansion for determining the fluxes $\mathcal{F}_E(u)$ and $\mathcal{F}_J(u)$, the two versions are only equivalent up to the PN truncation considered. We illustrate the use of the integral version of the maximum entropy conjecture using the PN approximation in Table~\ref{tab:chif-epsilonf} and Figs.~\ref{fig:entropy vs spin},~\ref{fig:remnant mass and spin}. In these applications, instead of the fluxes, we use the conservative PN dynamics encoded in Eq.~\eqref{eq:Mj}. However, this is equivalent to a calculation using the fluxes, given the standard assumption of flux balance used in PN calculations~\cite{Blanchet:2024PN} (see Sec.~II~E of~\cite{LeTiec:2011ab} for a discussion of the relation between the Bondi mass and the PN binding energy).

These results motivate a  ``{maximum entropy conjecture for black hole mergers},'' which can be stated as:
\begin{quote}
\emph{The merger results in the formation of a Kerr black hole which maximizes the entropy of the system subject to balance laws for the energy and angular momentum.}
\end{quote}

Putting the conjecture in other words, the physically realized remnant corresponds to the configuration for which the remnant's entropy is an extremum. This thermodynamic criterion offers a compelling heuristic for defining a final state that can be applied directly to PN inputs, without requiring NR simulations. Of course, the agreement with the NR results is what gives us confidence in proposing this criterion. In particular, for an equal-mass, nonspinning binary, the value of $\chi_f$ at the peak of the entropy approaches the spin value of the remnant predicted by NR simulations at higher PN orders, as seen in  Fig.~\ref{fig:entropy vs spin} and Table~\ref{tab:chif-epsilonf}. This is also close to the spin value $\chi_c \simeq 0.681$ at which the specific heat $C_J$ undergoes a phase transition~\cite{Davies:1977bgr}, which raises the question if there is a deeper thermodynamic reason for this coincidence.

To further test the maximum entropy conjecture, in Fig.~\ref{fig:remnant mass and spin} we plot, for different PN orders, the remnant spin $\chi_f$ (left panel) and mass $M_f$ (right panel) derived using Eq.~\eqref{eq:Mj}, as a function of the symmetric mass ratio $\nu$ of the binary. The dashed lines correspond to the predictions of the remnant mass and spin based on  the fits to NR simulations from Ref.~\cite{Jimenez-Forteza:2016oae}. For the displayed 1,000 non-spinning systems with mass ratio $q \in[1, 16]$, the remnant spin and mass at higher PN orders are in excellent agreement with the general relativistic prediction, accurate to within $5.6\%$ and $1.3\%$, respectively.

\emph{Results from NR simulations}---A central challenge in applying the maximum entropy conjecture to NR simulations is that there is no unique, universally accepted definition of the entropy of a binary system. 
Given this ambiguity, we defer an exact determination of the entropy for the BBH in general relativity to future work and adopt a pragmatic approach parallel to that used in the PN analysis—namely, mapping the system to a Kerr black hole and using $S_\mathrm{Kerr}$ in (\ref{eq:Sofj}). Given an NR simulation, we use the waveforms corrected for the center of mass drift, and extrapolated to future null infinity $\mathcal{I}^+$ to compute the energy and angular momentum flux of gravitational waves at each point in the binary's evolution [see, e.g., Eqs.~(1)--(4) in~\cite{Damour:2011fu}], starting from the simulation's reference time, after the ``junk'' radiation has left the domain. We then integrate these to obtain the energy and angular momentum of the binary at each point in its evolution. As in Refs.~\cite{Nagar:2015xqa,Ossokine:2017dge}, we apply a shift to the energy and angular momentum relation, so that the final point in the simulation agrees with the mass and angular momentum of the remnant black hole, as obtained from the corresponding apparent horizon (see~\cite{Boyle:2019kee}). This circumvents problems due to imperfect astrophysical initial data for the simulation.

\begin{figure*}
        \centering
        \includegraphics[width=0.48\linewidth]{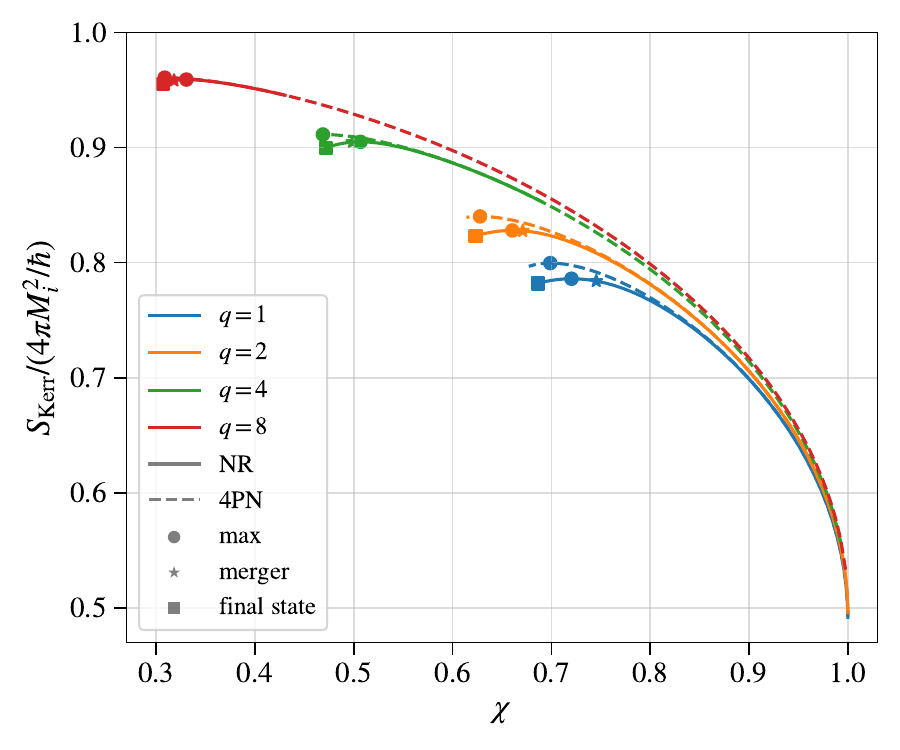}
        \includegraphics[width=0.50\linewidth]{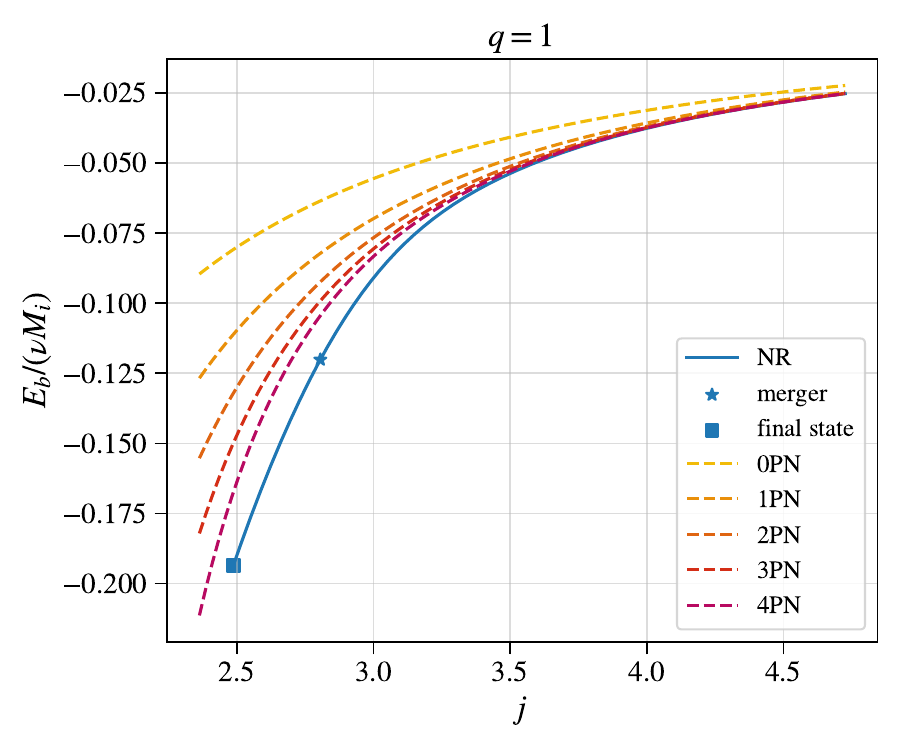}
        \caption{
        \emph{Left:} The entropy versus dimensionless spin of a Kerr black hole with the same energy and angular momentum as that at a particular point of a numerical relativity binary black hole simulation, along with the $4$PN prediction, all shown for nonspinning systems with mass ratios $q=M_1/M_2\ge 1$ of up to $8$ (SXS:BBH:3617, 1167, 2499, and 2707). The merger point corresponds to the peak of the amplitude of the dominant quadrupolar mode of the waveform, as in~\cite{Nagar:2015xqa}.
        \emph{Right:} The scaled binding energy versus angular momentum curve from the NR simulation for the equal-mass nonspinning case, with the merger and final state marked, compared with the PN predictions at the available orders.}
        \label{fig:SXS_S_vs_chi_nonspinning}
\end{figure*}

\begin{figure}[bht!]
    \centering
    \includegraphics[width=\linewidth]{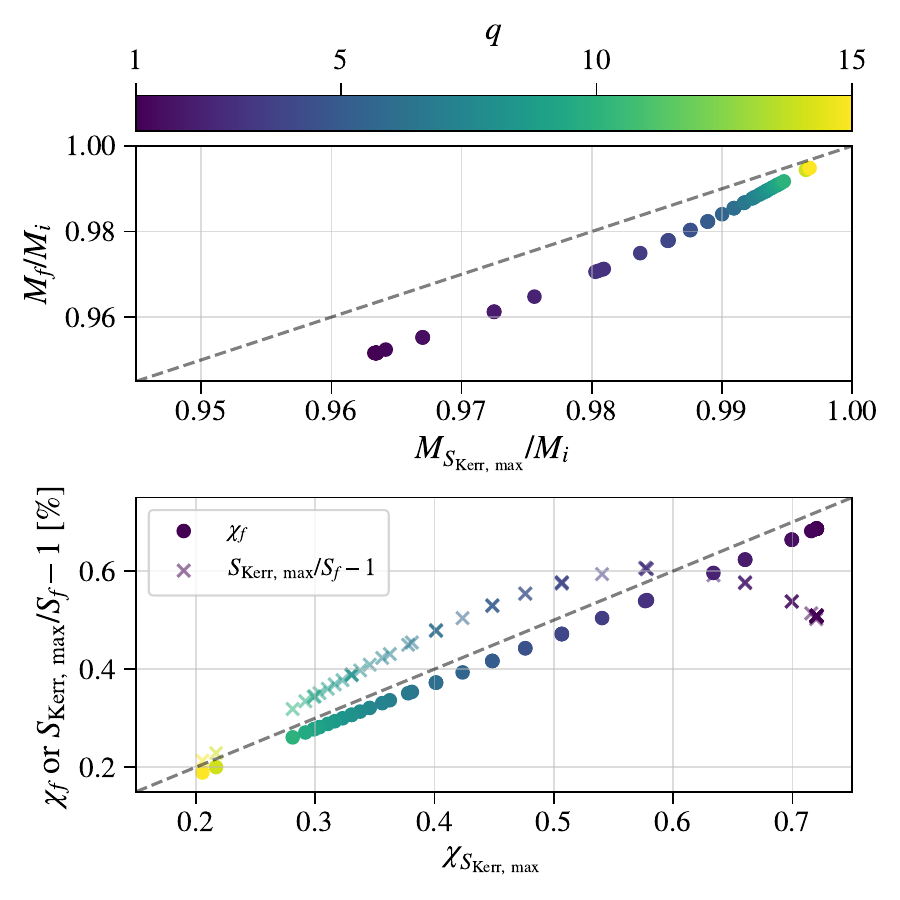}
    \caption{The final mass (top panel) and dimensionless spin (bottom panel) of the NR simulations versus the values one would obtain at the point where a Kerr black hole would have the maximum entropy, along with the identity map (dashed line). The bottom panel also shows the fractional difference in entropy between the maximum Kerr entropy and that of the final state. Colors give the binary's mass ratio.}
    \label{fig:Mf_chif_and_Sdiff_vs_Smax_vals}
\end{figure}

We restrict to NR simulations of non-spinning binaries with negligible eccentricity, and report results for binaries with aligned spins in a companion paper \cite{RinconRamirez:2025maxent}. We use the publicly available simulations from v3.0.0 of the SXS catalog~\cite{Scheel:2025jct},  always using the highest resolution simulation available. Additionally, we confirmed our results with the second highest resolution as described in a companion paper \cite{RinconRamirez:2025maxent}. 

For NR simulations, we use the same form of the entropy as in Eq.~\eqref{eq:Sofj}; however, the mass $M(j)$ and spin $J(j)$ are now given by their residual values at a retarded time $u$ on $\mathcal{I^+}$. This is the simplest, most straightforward model that can be conceived. We wish to point out that this entropic model is neither equivalent to the PN one, nor a unique choice that one can make. In the PN scenario, the mass and the spin of the system were modeled by the PN conservative dynamics. In contrast, in the NR case, it is modeled by the \emph{energetics}, which include not only the dynamics of the two black holes, but also the gravitational radiation in the spacetime. Additionally, as the dynamical horizon $\mathcal{H}$ is causally disconnected from $\mathcal{I}^+$ \cite{Ashtekar:2025wnu}, there is no fiducial map that relates the evolution of its area $A_{\mathcal{H}}(v)$ with that of the Bondi mass $M(u)$ and angular momentum $J(u)$. However, in practice, one can construct a map between the advanced time $v$ at $\mathcal{H}$ and the retarded time $u$ at $\mathcal{I}^+$ by cross-correlating the shear at $\mathcal{H}$ with the gravitational waves at $\mathcal{I}^+$ \cite{Jaramillo:2012rr,Chen:2022dxt,Prasad:2024vsz}.

The results are plotted in Fig.~\ref{fig:SXS_S_vs_chi_nonspinning} for four simulations with different mass ratios. The left panel shows the normalized entropy of a remnant if it were to form at a specific point in the simulation, versus the binary's energy and angular momentum at that point converted to the remnant dimensionless spin $\chi$ (the horizontal axis of the plot). In all cases, the predicted state from the maximum entropy prescription was slightly earlier in time than the actual final state in  the simulation, with the latter having a marginally lower entropy. The difference between the maximum entropy obtained with this method, and that of the final state is $0.61\%$ across all 62 nonspinning, quasi-circular simulations in the SXS catalog~\cite{Scheel:2025jct} that we analyzed, as illustrated in the bottom panel of Fig.~\ref{fig:Mf_chif_and_Sdiff_vs_Smax_vals}.

For non-spinning systems, across all mass ratios, the 4PN expressions predict a spin at maximum entropy that is closer to the final state than the NR-based maximum-Kerr-entropy estimate, while the corresponding mass shows a somewhat poorer agreement. This discrepancy is likely an artifact of the finite PN order (see Fig.~\ref{fig:SXS_S_vs_chi_nonspinning}). The mass and spin at maximum Kerr entropy remain close to, but slightly larger than, those of the final state (Fig.~\ref{fig:Mf_chif_and_Sdiff_vs_Smax_vals}).

% Finally, we can understand some of how the PN expressions can give results that are quantitatively in reasonably good agreement with NR even in regions where we would not \emph{a priori} expect them to be at all accurate by looking at the simplicity of the binding energy $E_b$ versus $j$ curves. (as in~\cite{Damour:2011fu,Nagar:2015xqa,Ossokine:2017dge} we scale the binding energy by the binary's reduced mass). We illustrate this for the equal-mass non-spinning case in Fig.~\ref{fig:SXS_S_vs_chi_nonspinning}, right panel, showing that the NR curve remains concave down with no noticeable change in its behavior all the way to the final state. The $1/j^2$ expansion parameter [see, e.g., Eq.~\eqref{eq:Mj}] is also not particularly large for the smallest values of $j$ encountered in the non-spinning case ($j \simeq 2.5$ $\Rightarrow$ $1/j^2 \simeq 0.16$, and the unknown $5$PN contribution involves a factor of $1/j^{12} \simeq 2 \times 10^{-5}$).  Additionally, while the $4$PN binding energy as a function of orbital angular momentum agrees reasonably well with the NR result up to merger within a maximum fractional difference of $15\%$, the fractional difference in the entropy is better than $2\%$. 

Finally, the surprisingly good agreement between the PN expressions and NR results, even close to merger, can be understood qualitatively from the simplicity of the NR binding-energy curve $E_b(j)$. As shown in Fig.~\ref{fig:SXS_S_vs_chi_nonspinning} (right; see~\cite{Nagar:2015xqa, Ossokine:2017dge} for similar plots for other binaries), the equal-mass, non-spinning NR curve remains smooth and concave down all the way to the final state, with no sudden change in behavior that would strongly penalize a truncated expansion, even though the energy and angular momentum are computed from the gravitational waveform, which has a significant change in morphology at merger (see~\cite{Scheel:2008rj}). Moreover, the PN expansion parameter $1/j^2$ is still modest in the regime of interest (e.g., $j \simeq 2.5 \Rightarrow 1/j^2 \simeq 0.16$), and the unknown $5$PN contribution involves a factor of $1/j^{12} \lesssim 2 \times 10^{-5}$, suggesting that the neglected higher-order contributions, while not rigorously controlled, are unlikely to become parametrically large near merger. In fact, for an equal-mass nonspinning binary, the $4$PN binding energy as a function of orbital angular momentum agrees reasonably well with the NR result up to merger, with a maximum fractional difference of $15\%$. The Kerr entropy obtained from the $4$PN energy-angular momentum relation agrees even better with NR, with a maximum fractional difference of $2\%.$

The results from both PN theory and NR simulations show that the maximum-entropy point reproduces the remnant’s mass and spin to within a few percent of the true values. This suggests that the endpoint of a merger is thermodynamically determined, revealing a new deep connection between black-hole dynamics and the laws of thermodynamics.

\begin{acknowledgments}
We thank Abhay Ashtekar, Luc Blanchet, Koustav Chandra, Jolien Creighton, Hal Haggard, Luis Lehner, and Harald Pfeiffer for insightful discussions and comments that led to improvements in the clarity and presentation of the paper. N.K.J.-M.\ acknowledges support from NSF grant AST-2205920,  I.G. from PHY-2020275 (Network for Neutrinos, Nuclear Astrophysics, and Symmetries (N3AS)), E.B.  from PHY-2207851, PHY-2513194, and ID 63683 grant from the John Templeton Foundation, as part of the ``WithOut SpaceTime'' (\href{https://withoutspacetime.org/}{WOST}) consortium, and B.S.S. from AST-2307147, PHY-2207638, PHY-2308886 and PHY-2309064. This work used the software packages \texttt{LALSuite}~\cite{LALSuite}, \texttt{matplotlib}~\cite{Hunter:2007ouj}, \texttt{numpy}~\cite{Harris:2020xlr}, and \texttt{sxs}~\cite{sxs_package}. \end{acknowledgments}

\bibliography{mybib}

\providecommand{\noopsort}[1]{}\providecommand{\singleletter}[1]{#1}%
\begin{thebibliography}{45}%
\makeatletter
\providecommand \@ifxundefined [1]{%
 \@ifx{#1\undefined}
}%
\providecommand \@ifnum [1]{%
 \ifnum #1\expandafter \@firstoftwo
 \else \expandafter \@secondoftwo
 \fi
}%
\providecommand \@ifx [1]{%
 \ifx #1\expandafter \@firstoftwo
 \else \expandafter \@secondoftwo
 \fi
}%
\providecommand \natexlab [1]{#1}%
\providecommand \enquote  [1]{``#1''}%
\providecommand \bibnamefont  [1]{#1}%
\providecommand \bibfnamefont [1]{#1}%
\providecommand \citenamefont [1]{#1}%
\providecommand \href@noop [0]{\@secondoftwo}%
\providecommand \href [0]{\begingroup \@sanitize@url \@href}%
\providecommand \@href[1]{\@@startlink{#1}\@@href}%
\providecommand \@@href[1]{\endgroup#1\@@endlink}%
\providecommand \@sanitize@url [0]{\catcode `\\12\catcode `\$12\catcode
  `\&12\catcode `\#12\catcode `\^12\catcode `\_12\catcode `\%12\relax}%
\providecommand \@@startlink[1]{}%
\providecommand \@@endlink[0]{}%
\providecommand \url  [0]{\begingroup\@sanitize@url \@url }%
\providecommand \@url [1]{\endgroup\@href {#1}{\urlprefix }}%
\providecommand \urlprefix  [0]{URL }%
\providecommand \Eprint [0]{\href }%
\providecommand \doibase [0]{https://doi.org/}%
\providecommand \selectlanguage [0]{\@gobble}%
\providecommand \bibinfo  [0]{\@secondoftwo}%
\providecommand \bibfield  [0]{\@secondoftwo}%
\providecommand \translation [1]{[#1]}%
\providecommand \BibitemOpen [0]{}%
\providecommand \bibitemStop [0]{}%
\providecommand \bibitemNoStop [0]{.\EOS\space}%
\providecommand \EOS [0]{\spacefactor3000\relax}%
\providecommand \BibitemShut  [1]{\csname bibitem#1\endcsname}%
\let\auto@bib@innerbib\@empty
%</preamble>
\bibitem [{\citenamefont {Abac}\ \emph
  {et~al.}(2025{\natexlab{a}})\citenamefont {Abac} \emph
  {et~al.}}]{LIGOScientific:2025slb}%
  \BibitemOpen
  \bibfield  {author} {\bibinfo {author} {\bibfnamefont {A.~G.}\ \bibnamefont
  {Abac}} \emph {et~al.} (\bibinfo {collaboration} {LIGO Scientific, Virgo, and
  KAGRA Collaborations}),\ }\href@noop {} {\bibinfo {title} {{GWTC-4.0:
  Updating the Gravitational-Wave Transient Catalog with Observations from the
  First Part of the Fourth LIGO-Virgo-KAGRA Observing Run}}} (\bibinfo {year}
  {2025}{\natexlab{a}}),\ \Eprint {https://arxiv.org/abs/2508.18082}
  {arXiv:2508.18082 [gr-qc]} \BibitemShut {NoStop}%
\bibitem [{\citenamefont {Baumgarte}\ and\ \citenamefont
  {Shapiro}(2010)}]{Baumgarte:2010ndz}%
  \BibitemOpen
  \bibfield  {author} {\bibinfo {author} {\bibfnamefont {T.~W.}\ \bibnamefont
  {Baumgarte}}\ and\ \bibinfo {author} {\bibfnamefont {S.~L.}\ \bibnamefont
  {Shapiro}},\ }\href {https://doi.org/10.1017/CBO9781139193344} {\emph
  {\bibinfo {title} {{Numerical Relativity: Solving Einstein's Equations on the
  Computer}}}}\ (\bibinfo  {publisher} {Cambridge University Press},\ \bibinfo
  {year} {2010})\BibitemShut {NoStop}%
\bibitem [{\citenamefont {Scheel}\ \emph {et~al.}(2025)\citenamefont {Scheel}
  \emph {et~al.}}]{Scheel:2025jct}%
  \BibitemOpen
  \bibfield  {author} {\bibinfo {author} {\bibfnamefont {M.~A.}\ \bibnamefont
  {Scheel}} \emph {et~al.},\ }\bibfield  {title} {\bibinfo {title} {{The SXS
  collaboration{\textquoteright}s third catalog of binary black hole
  simulations}},\ }\href {https://doi.org/10.1088/1361-6382/adfd34} {\bibfield
  {journal} {\bibinfo  {journal} {Classical Quantum Gravity}\ }\textbf
  {\bibinfo {volume} {42}},\ \bibinfo {pages} {195017} (\bibinfo {year}
  {2025})},\ \Eprint {https://arxiv.org/abs/2505.13378} {arXiv:2505.13378
  [gr-qc]} \BibitemShut {NoStop}%
\bibitem [{\citenamefont {Scheel}\ \emph {et~al.}(2009)\citenamefont {Scheel},
  \citenamefont {Boyle}, \citenamefont {Chu}, \citenamefont {Kidder},
  \citenamefont {Matthews},\ and\ \citenamefont {Pfeiffer}}]{Scheel:2008rj}%
  \BibitemOpen
  \bibfield  {author} {\bibinfo {author} {\bibfnamefont {M.~A.}\ \bibnamefont
  {Scheel}}, \bibinfo {author} {\bibfnamefont {M.}~\bibnamefont {Boyle}},
  \bibinfo {author} {\bibfnamefont {T.}~\bibnamefont {Chu}}, \bibinfo {author}
  {\bibfnamefont {L.~E.}\ \bibnamefont {Kidder}}, \bibinfo {author}
  {\bibfnamefont {K.~D.}\ \bibnamefont {Matthews}},\ and\ \bibinfo {author}
  {\bibfnamefont {H.~P.}\ \bibnamefont {Pfeiffer}},\ }\bibfield  {title}
  {\bibinfo {title} {{High-accuracy waveforms for binary black hole inspiral,
  merger, and ringdown}},\ }\href {https://doi.org/10.1103/PhysRevD.79.024003}
  {\bibfield  {journal} {\bibinfo  {journal} {Phys. Rev. D}\ }\textbf {\bibinfo
  {volume} {79}},\ \bibinfo {pages} {024003} (\bibinfo {year} {2009})},\
  \Eprint {https://arxiv.org/abs/0810.1767} {arXiv:0810.1767 [gr-qc]}
  \BibitemShut {NoStop}%
\bibitem [{Note1()}]{Note1}%
  \BibitemOpen
  \bibinfo {note} {We work in units with $c=1,$ $G=1,$ and $k_B=1$, keeping
  track of Planck's constant $\hbar $ to distinguish quantum from classical
  phenomena.}\BibitemShut {Stop}%
\bibitem [{\citenamefont {Buonanno}\ \emph {et~al.}(2008)\citenamefont
  {Buonanno}, \citenamefont {Kidder},\ and\ \citenamefont
  {Lehner}}]{Buonanno:2007sv}%
  \BibitemOpen
  \bibfield  {author} {\bibinfo {author} {\bibfnamefont {A.}~\bibnamefont
  {Buonanno}}, \bibinfo {author} {\bibfnamefont {L.~E.}\ \bibnamefont
  {Kidder}},\ and\ \bibinfo {author} {\bibfnamefont {L.}~\bibnamefont
  {Lehner}},\ }\bibfield  {title} {\bibinfo {title} {{Estimating the final spin
  of a binary black hole coalescence}},\ }\href
  {https://doi.org/10.1103/PhysRevD.77.026004} {\bibfield  {journal} {\bibinfo
  {journal} {Phys. Rev. D}\ }\textbf {\bibinfo {volume} {77}},\ \bibinfo
  {pages} {026004} (\bibinfo {year} {2008})},\ \Eprint
  {https://arxiv.org/abs/0709.3839} {arXiv:0709.3839 [astro-ph]} \BibitemShut
  {NoStop}%
\bibitem [{\citenamefont {Abbott}\ \emph {et~al.}(2016)\citenamefont {Abbott}
  \emph {et~al.}}]{LIGOScientific:2016vlm}%
  \BibitemOpen
  \bibfield  {author} {\bibinfo {author} {\bibfnamefont {B.~P.}\ \bibnamefont
  {Abbott}} \emph {et~al.} (\bibinfo {collaboration} {LIGO Scientific and Virgo
  Collaborations}),\ }\bibfield  {title} {\bibinfo {title} {{Properties of the
  Binary Black Hole Merger GW150914}},\ }\href
  {https://doi.org/10.1103/PhysRevLett.116.241102} {\bibfield  {journal}
  {\bibinfo  {journal} {Phys. Rev. Lett.}\ }\textbf {\bibinfo {volume} {116}},\
  \bibinfo {pages} {241102} (\bibinfo {year} {2016})},\ \Eprint
  {https://arxiv.org/abs/1602.03840} {arXiv:1602.03840 [gr-qc]} \BibitemShut
  {NoStop}%
\bibitem [{\citenamefont {Davies}(1977)}]{Davies:1977bgr}%
  \BibitemOpen
  \bibfield  {author} {\bibinfo {author} {\bibfnamefont {P.~C.~W.}\
  \bibnamefont {Davies}},\ }\bibfield  {title} {\bibinfo {title}
  {{Thermodynamics of Black Holes}},\ }\href
  {https://doi.org/10.1098/rspa.1977.0047} {\bibfield  {journal} {\bibinfo
  {journal} {Proc. R. Soc. A}\ }\textbf {\bibinfo {volume} {353}},\ \bibinfo
  {pages} {499} (\bibinfo {year} {1977})}\BibitemShut {NoStop}%
\bibitem [{\citenamefont {Hofmann}\ \emph {et~al.}(2016)\citenamefont
  {Hofmann}, \citenamefont {Barausse},\ and\ \citenamefont
  {Rezzolla}}]{Hofmann:2016yih}%
  \BibitemOpen
  \bibfield  {author} {\bibinfo {author} {\bibfnamefont {F.}~\bibnamefont
  {Hofmann}}, \bibinfo {author} {\bibfnamefont {E.}~\bibnamefont {Barausse}},\
  and\ \bibinfo {author} {\bibfnamefont {L.}~\bibnamefont {Rezzolla}},\
  }\bibfield  {title} {\bibinfo {title} {{The final spin from binary black
  holes in quasi-circular orbits}},\ }\href
  {https://doi.org/10.3847/2041-8205/825/2/L19} {\bibfield  {journal} {\bibinfo
   {journal} {Astrophys. J. Lett.}\ }\textbf {\bibinfo {volume} {825}},\
  \bibinfo {pages} {L19} (\bibinfo {year} {2016})},\ \Eprint
  {https://arxiv.org/abs/1605.01938} {arXiv:1605.01938 [gr-qc]} \BibitemShut
  {NoStop}%
\bibitem [{\citenamefont {Healy}\ and\ \citenamefont
  {Lousto}(2017)}]{Healy:2016lce}%
  \BibitemOpen
  \bibfield  {author} {\bibinfo {author} {\bibfnamefont {J.}~\bibnamefont
  {Healy}}\ and\ \bibinfo {author} {\bibfnamefont {C.~O.}\ \bibnamefont
  {Lousto}},\ }\bibfield  {title} {\bibinfo {title} {{Remnant of binary
  black-hole mergers: New simulations and peak luminosity studies}},\ }\href
  {https://doi.org/10.1103/PhysRevD.95.024037} {\bibfield  {journal} {\bibinfo
  {journal} {Phys. Rev. D}\ }\textbf {\bibinfo {volume} {95}},\ \bibinfo
  {pages} {024037} (\bibinfo {year} {2017})},\ \Eprint
  {https://arxiv.org/abs/1610.09713} {arXiv:1610.09713 [gr-qc]} \BibitemShut
  {NoStop}%
%%CITATION = ARXIV:1610.09713;%%
\bibitem [{\citenamefont {Jim\'enez-Forteza}\ \emph {et~al.}(2017)\citenamefont
  {Jim\'enez-Forteza}, \citenamefont {Keitel}, \citenamefont {Husa},
  \citenamefont {Hannam}, \citenamefont {Khan},\ and\ \citenamefont
  {P\"urrer}}]{Jimenez-Forteza:2016oae}%
  \BibitemOpen
  \bibfield  {author} {\bibinfo {author} {\bibfnamefont {X.}~\bibnamefont
  {Jim\'enez-Forteza}}, \bibinfo {author} {\bibfnamefont {D.}~\bibnamefont
  {Keitel}}, \bibinfo {author} {\bibfnamefont {S.}~\bibnamefont {Husa}},
  \bibinfo {author} {\bibfnamefont {M.}~\bibnamefont {Hannam}}, \bibinfo
  {author} {\bibfnamefont {S.}~\bibnamefont {Khan}},\ and\ \bibinfo {author}
  {\bibfnamefont {M.}~\bibnamefont {P\"urrer}},\ }\bibfield  {title} {\bibinfo
  {title} {{Hierarchical data-driven approach to fitting numerical relativity
  data for nonprecessing binary black holes with an application to final spin
  and radiated energy}},\ }\href {https://doi.org/10.1103/PhysRevD.95.064024}
  {\bibfield  {journal} {\bibinfo  {journal} {Phys. Rev. D}\ }\textbf {\bibinfo
  {volume} {95}},\ \bibinfo {pages} {064024} (\bibinfo {year} {2017})},\
  \Eprint {https://arxiv.org/abs/1611.00332} {arXiv:1611.00332 [gr-qc]}
  \BibitemShut {NoStop}%
\bibitem [{\citenamefont {Varma}\ \emph {et~al.}(2019)\citenamefont {Varma},
  \citenamefont {Field}, \citenamefont {Scheel}, \citenamefont {Blackman},
  \citenamefont {Gerosa}, \citenamefont {Stein}, \citenamefont {Kidder},\ and\
  \citenamefont {Pfeiffer}}]{PhysRevResearch.1.033015}%
  \BibitemOpen
  \bibfield  {author} {\bibinfo {author} {\bibfnamefont {V.}~\bibnamefont
  {Varma}}, \bibinfo {author} {\bibfnamefont {S.~E.}\ \bibnamefont {Field}},
  \bibinfo {author} {\bibfnamefont {M.~A.}\ \bibnamefont {Scheel}}, \bibinfo
  {author} {\bibfnamefont {J.}~\bibnamefont {Blackman}}, \bibinfo {author}
  {\bibfnamefont {D.}~\bibnamefont {Gerosa}}, \bibinfo {author} {\bibfnamefont
  {L.~C.}\ \bibnamefont {Stein}}, \bibinfo {author} {\bibfnamefont {L.~E.}\
  \bibnamefont {Kidder}},\ and\ \bibinfo {author} {\bibfnamefont {H.~P.}\
  \bibnamefont {Pfeiffer}},\ }\bibfield  {title} {\bibinfo {title} {Surrogate
  models for precessing binary black hole simulations with unequal masses},\
  }\href {https://doi.org/10.1103/PhysRevResearch.1.033015} {\bibfield
  {journal} {\bibinfo  {journal} {Phys. Rev. Research}\ }\textbf {\bibinfo
  {volume} {1}},\ \bibinfo {pages} {033015} (\bibinfo {year}
  {2019})}\BibitemShut {NoStop}%
\bibitem [{\citenamefont {Bardeen}\ \emph {et~al.}(1973)\citenamefont
  {Bardeen}, \citenamefont {Carter},\ and\ \citenamefont
  {Hawking}}]{Bardeen:1973gs}%
  \BibitemOpen
  \bibfield  {author} {\bibinfo {author} {\bibfnamefont {J.~M.}\ \bibnamefont
  {Bardeen}}, \bibinfo {author} {\bibfnamefont {B.}~\bibnamefont {Carter}},\
  and\ \bibinfo {author} {\bibfnamefont {S.~W.}\ \bibnamefont {Hawking}},\
  }\bibfield  {title} {\bibinfo {title} {{The Four laws of black hole
  mechanics}},\ }\href {https://doi.org/10.1007/BF01645742} {\bibfield
  {journal} {\bibinfo  {journal} {Commun. Math. Phys.}\ }\textbf {\bibinfo
  {volume} {31}},\ \bibinfo {pages} {161} (\bibinfo {year} {1973})}\BibitemShut
  {NoStop}%
\bibitem [{\citenamefont {Bekenstein}(1973)}]{Bekenstein:1973ur}%
  \BibitemOpen
  \bibfield  {author} {\bibinfo {author} {\bibfnamefont {J.~D.}\ \bibnamefont
  {Bekenstein}},\ }\bibfield  {title} {\bibinfo {title} {Black holes and
  entropy},\ }\href {https://doi.org/10.1103/PhysRevD.7.2333} {\bibfield
  {journal} {\bibinfo  {journal} {Phys. Rev. D}\ }\textbf {\bibinfo {volume}
  {7}},\ \bibinfo {pages} {2333} (\bibinfo {year} {1973})}\BibitemShut
  {NoStop}%
\bibitem [{\citenamefont {Hawking}(1971)}]{Hawking:1971tu}%
  \BibitemOpen
  \bibfield  {author} {\bibinfo {author} {\bibfnamefont {S.~W.}\ \bibnamefont
  {Hawking}},\ }\bibfield  {title} {\bibinfo {title} {{Gravitational radiation
  from colliding black holes}},\ }\href
  {https://doi.org/10.1103/PhysRevLett.26.1344} {\bibfield  {journal} {\bibinfo
   {journal} {Phys. Rev. Lett.}\ }\textbf {\bibinfo {volume} {26}},\ \bibinfo
  {pages} {1344} (\bibinfo {year} {1971})}\BibitemShut {NoStop}%
\bibitem [{\citenamefont {Ashtekar}\ and\ \citenamefont
  {Krishnan}(2025)}]{Ashtekar:2025wnu}%
  \BibitemOpen
  \bibfield  {author} {\bibinfo {author} {\bibfnamefont {A.}~\bibnamefont
  {Ashtekar}}\ and\ \bibinfo {author} {\bibfnamefont {B.}~\bibnamefont
  {Krishnan}},\ }\bibfield  {title} {\bibinfo {title} {{Quasi-local black hole
  horizons: recent advances}},\ }\href
  {https://doi.org/10.1007/s41114-025-00061-4} {\bibfield  {journal} {\bibinfo
  {journal} {Living Rev. Relativity}\ }\textbf {\bibinfo {volume} {28}},\
  \bibinfo {pages} {8} (\bibinfo {year} {2025})},\ \Eprint
  {https://arxiv.org/abs/2502.11825} {arXiv:2502.11825 [gr-qc]} \BibitemShut
  {NoStop}%
\bibitem [{\citenamefont {Abac}\ \emph
  {et~al.}(2025{\natexlab{b}})\citenamefont {Abac} \emph
  {et~al.}}]{LIGOScientific:2025rid}%
  \BibitemOpen
  \bibfield  {author} {\bibinfo {author} {\bibfnamefont {A.~G.}\ \bibnamefont
  {Abac}} \emph {et~al.} (\bibinfo {collaboration} {LIGO Scientific, Virgo, and
  KAGRA Collaborations}),\ }\bibfield  {title} {\bibinfo {title} {{GW250114:
  Testing Hawking{\textquoteright}s Area Law and the Kerr Nature of Black
  Holes}},\ }\href {https://doi.org/10.1103/kw5g-d732} {\bibfield  {journal}
  {\bibinfo  {journal} {Phys. Rev. Lett.}\ }\textbf {\bibinfo {volume} {135}},\
  \bibinfo {pages} {111403} (\bibinfo {year} {2025}{\natexlab{b}})},\ \Eprint
  {https://arxiv.org/abs/2509.08054} {arXiv:2509.08054 [gr-qc]} \BibitemShut
  {NoStop}%
\bibitem [{\citenamefont {Fermi}(1956)}]{fermi1956thermodynamics}%
  \BibitemOpen
  \bibfield  {author} {\bibinfo {author} {\bibfnamefont {E.}~\bibnamefont
  {Fermi}},\ }\href
  {https://store.doverpublications.com/products/9780486603612} {\emph {\bibinfo
  {title} {Thermodynamics}}}\ (\bibinfo  {publisher} {Dover Publications},\
  \bibinfo {year} {1956})\ \bibinfo {note} {reprint of lectures delivered at
  Columbia University}\BibitemShut {NoStop}%
\bibitem [{\citenamefont {{Callen}}(1991)}]{callen1985thermodynamics}%
  \BibitemOpen
  \bibfield  {author} {\bibinfo {author} {\bibfnamefont {H.~B.}\ \bibnamefont
  {{Callen}}},\ }\href
  {https://www.wiley.com/en-us/Thermodynamics+and+an+Introduction+to+Thermostatistics%2C+2nd+Edition-p-9780471862567}
  {\emph {\bibinfo {title} {{Thermodynamics and an Introduction to
  Thermostatistics}}}},\ \bibinfo {edition} {2nd}\ ed.\ (\bibinfo  {publisher}
  {Wiley},\ \bibinfo {year} {1991})\BibitemShut {NoStop}%
\bibitem [{\citenamefont {Blanchet}(2024)}]{Blanchet:2024PN}%
  \BibitemOpen
  \bibfield  {author} {\bibinfo {author} {\bibfnamefont {L.}~\bibnamefont
  {Blanchet}},\ }\bibfield  {title} {\bibinfo {title} {{Post-Newtonian theory
  for gravitational waves}},\ }\href
  {https://doi.org/10.1007/s41114-024-00050-z} {\bibfield  {journal} {\bibinfo
  {journal} {Living Rev. Relativity}\ }\textbf {\bibinfo {volume} {27}},\
  \bibinfo {pages} {4} (\bibinfo {year} {2024})}\BibitemShut {NoStop}%
\bibitem [{\citenamefont {Blanchet}\ \emph {et~al.}(2025)\citenamefont
  {Blanchet}, \citenamefont {Langlois},\ and\ \citenamefont
  {Ligout}}]{Blanchet:2025agj}%
  \BibitemOpen
  \bibfield  {author} {\bibinfo {author} {\bibfnamefont {L.}~\bibnamefont
  {Blanchet}}, \bibinfo {author} {\bibfnamefont {D.}~\bibnamefont {Langlois}},\
  and\ \bibinfo {author} {\bibfnamefont {E.}~\bibnamefont {Ligout}},\
  }\bibfield  {title} {\bibinfo {title} {{Innermost stable circular orbit of
  arbitrary-mass compact binaries at fourth post-Newtonian order}},\ }\href
  {https://doi.org/10.1103/mtv7-lkv8} {\bibfield  {journal} {\bibinfo
  {journal} {Phys. Rev. D}\ }\textbf {\bibinfo {volume} {112}},\ \bibinfo
  {pages} {064025} (\bibinfo {year} {2025})},\ \Eprint
  {https://arxiv.org/abs/2505.01278} {arXiv:2505.01278 [gr-qc]} \BibitemShut
  {NoStop}%
\bibitem [{\citenamefont {Cabero}\ \emph {et~al.}(2017)\citenamefont {Cabero},
  \citenamefont {Nielsen}, \citenamefont {Lundgren},\ and\ \citenamefont
  {Capano}}]{Cabero:2016ayq}%
  \BibitemOpen
  \bibfield  {author} {\bibinfo {author} {\bibfnamefont {M.}~\bibnamefont
  {Cabero}}, \bibinfo {author} {\bibfnamefont {A.~B.}\ \bibnamefont {Nielsen}},
  \bibinfo {author} {\bibfnamefont {A.~P.}\ \bibnamefont {Lundgren}},\ and\
  \bibinfo {author} {\bibfnamefont {C.~D.}\ \bibnamefont {Capano}},\ }\bibfield
   {title} {\bibinfo {title} {{Minimum energy and the end of the inspiral in
  the post-Newtonian approximation}},\ }\href
  {https://doi.org/10.1103/PhysRevD.95.064016} {\bibfield  {journal} {\bibinfo
  {journal} {Phys. Rev. D}\ }\textbf {\bibinfo {volume} {95}},\ \bibinfo
  {pages} {064016} (\bibinfo {year} {2017})},\ \Eprint
  {https://arxiv.org/abs/1602.03134} {arXiv:1602.03134 [gr-qc]} \BibitemShut
  {NoStop}%
\bibitem [{Note2()}]{Note2}%
  \BibitemOpen
  \bibinfo {note} {It is important to note that the post-Newtonian expansion is
  an asymptotic series that is not guaranteed to monotonically approach its
  limiting value \cite {Blanchet:2024PN, Damour:1997ub}.}\BibitemShut {Stop}%
\bibitem [{\citenamefont {Damour}\ \emph {et~al.}(2014)\citenamefont {Damour},
  \citenamefont {Jaranowski},\ and\ \citenamefont {Sch\"afer}}]{4PN_Damour}%
  \BibitemOpen
  \bibfield  {author} {\bibinfo {author} {\bibfnamefont {T.}~\bibnamefont
  {Damour}}, \bibinfo {author} {\bibfnamefont {P.}~\bibnamefont {Jaranowski}},\
  and\ \bibinfo {author} {\bibfnamefont {G.}~\bibnamefont {Sch\"afer}},\
  }\bibfield  {title} {\bibinfo {title} {{Nonlocal-in-time action for the
  fourth post-Newtonian conservative dynamics of two-body systems}},\ }\href
  {https://doi.org/10.1103/PhysRevD.89.064058} {\bibfield  {journal} {\bibinfo
  {journal} {Phys. Rev. D}\ }\textbf {\bibinfo {volume} {89}},\ \bibinfo
  {pages} {064058} (\bibinfo {year} {2014})}\BibitemShut {NoStop}%
\bibitem [{\citenamefont {Kerr}(1963)}]{Kerr:1963ud}%
  \BibitemOpen
  \bibfield  {author} {\bibinfo {author} {\bibfnamefont {R.~P.}\ \bibnamefont
  {Kerr}},\ }\bibfield  {title} {\bibinfo {title} {{Gravitational field of a
  spinning mass as an example of algebraically special metrics}},\ }\href
  {https://doi.org/10.1103/PhysRevLett.11.237} {\bibfield  {journal} {\bibinfo
  {journal} {Phys. Rev. Lett.}\ }\textbf {\bibinfo {volume} {11}},\ \bibinfo
  {pages} {237} (\bibinfo {year} {1963})}\BibitemShut {NoStop}%
\bibitem [{\citenamefont {Ashtekar}\ and\ \citenamefont
  {Krishnan}(2004)}]{Ashtekar:2004cn}%
  \BibitemOpen
  \bibfield  {author} {\bibinfo {author} {\bibfnamefont {A.}~\bibnamefont
  {Ashtekar}}\ and\ \bibinfo {author} {\bibfnamefont {B.}~\bibnamefont
  {Krishnan}},\ }\bibfield  {title} {\bibinfo {title} {{Isolated and dynamical
  horizons and their applications}},\ }\href
  {https://doi.org/10.12942/lrr-2004-10} {\bibfield  {journal} {\bibinfo
  {journal} {Living Rev. Rel.}\ }\textbf {\bibinfo {volume} {7}},\ \bibinfo
  {pages} {10} (\bibinfo {year} {2004})},\ \Eprint
  {https://arxiv.org/abs/gr-qc/0407042} {arXiv:gr-qc/0407042} \BibitemShut
  {NoStop}%
\bibitem [{\citenamefont {Scheel}\ \emph {et~al.}(2015)\citenamefont {Scheel},
  \citenamefont {Giesler}, \citenamefont {Hemberger}, \citenamefont {Lovelace},
  \citenamefont {Kuper}, \citenamefont {Boyle}, \citenamefont {Szil{\'a}gyi},\
  and\ \citenamefont {Kidder}}]{Scheel:2014ina}%
  \BibitemOpen
  \bibfield  {author} {\bibinfo {author} {\bibfnamefont {M.~A.}\ \bibnamefont
  {Scheel}}, \bibinfo {author} {\bibfnamefont {M.}~\bibnamefont {Giesler}},
  \bibinfo {author} {\bibfnamefont {D.~A.}\ \bibnamefont {Hemberger}}, \bibinfo
  {author} {\bibfnamefont {G.}~\bibnamefont {Lovelace}}, \bibinfo {author}
  {\bibfnamefont {K.}~\bibnamefont {Kuper}}, \bibinfo {author} {\bibfnamefont
  {M.}~\bibnamefont {Boyle}}, \bibinfo {author} {\bibfnamefont
  {B.}~\bibnamefont {Szil{\'a}gyi}},\ and\ \bibinfo {author} {\bibfnamefont
  {L.~E.}\ \bibnamefont {Kidder}},\ }\bibfield  {title} {\bibinfo {title}
  {{Improved methods for simulating nearly extremal binary black holes}},\
  }\href {https://doi.org/10.1088/0264-9381/32/10/105009} {\bibfield  {journal}
  {\bibinfo  {journal} {Classical Quantum Gravity}\ }\textbf {\bibinfo {volume}
  {32}},\ \bibinfo {pages} {105009} (\bibinfo {year} {2015})},\ \Eprint
  {https://arxiv.org/abs/1412.1803} {arXiv:1412.1803 [gr-qc]} \BibitemShut
  {NoStop}%
\bibitem [{\citenamefont {Gupta}\ \emph {et~al.}(2018)\citenamefont {Gupta},
  \citenamefont {Krishnan}, \citenamefont {Nielsen},\ and\ \citenamefont
  {Schnetter}}]{Gupta:2018znn}%
  \BibitemOpen
  \bibfield  {author} {\bibinfo {author} {\bibfnamefont {A.}~\bibnamefont
  {Gupta}}, \bibinfo {author} {\bibfnamefont {B.}~\bibnamefont {Krishnan}},
  \bibinfo {author} {\bibfnamefont {A.~B.}\ \bibnamefont {Nielsen}},\ and\
  \bibinfo {author} {\bibfnamefont {E.}~\bibnamefont {Schnetter}},\ }\bibfield
  {title} {\bibinfo {title} {{Dynamics of marginally trapped surfaces in a
  binary black hole merger: Growth and approach to equilibrium}},\ }\href
  {https://doi.org/10.1103/PhysRevD.97.084028} {\bibfield  {journal} {\bibinfo
  {journal} {Phys. Rev. D}\ }\textbf {\bibinfo {volume} {97}},\ \bibinfo
  {pages} {084028} (\bibinfo {year} {2018})},\ \Eprint
  {https://arxiv.org/abs/1801.07048} {arXiv:1801.07048 [gr-qc]} \BibitemShut
  {NoStop}%
\bibitem [{\citenamefont {M{\"a}dler}\ and\ \citenamefont
  {Winicour}(2016)}]{M_dler_2016}%
  \BibitemOpen
  \bibfield  {author} {\bibinfo {author} {\bibfnamefont {T.}~\bibnamefont
  {M{\"a}dler}}\ and\ \bibinfo {author} {\bibfnamefont {J.}~\bibnamefont
  {Winicour}},\ }\bibfield  {title} {\bibinfo {title} {{Bondi-Sachs
  Formalism}},\ }\href {https://doi.org/10.4249/scholarpedia.33528} {\bibfield
  {journal} {\bibinfo  {journal} {Scholarpedia}\ }\textbf {\bibinfo {volume}
  {11}},\ \bibinfo {pages} {33528} (\bibinfo {year} {2016})}\BibitemShut
  {NoStop}%
\bibitem [{\citenamefont {Friedman}\ \emph {et~al.}(2002)\citenamefont
  {Friedman}, \citenamefont {Uryu},\ and\ \citenamefont
  {Shibata}}]{Friedman:2001pf}%
  \BibitemOpen
  \bibfield  {author} {\bibinfo {author} {\bibfnamefont {J.~L.}\ \bibnamefont
  {Friedman}}, \bibinfo {author} {\bibfnamefont {K.}~\bibnamefont {Uryu}},\
  and\ \bibinfo {author} {\bibfnamefont {M.}~\bibnamefont {Shibata}},\
  }\bibfield  {title} {\bibinfo {title} {{Thermodynamics of binary black holes
  and neutron stars}},\ }\href {https://doi.org/10.1103/PhysRevD.70.129904}
  {\bibfield  {journal} {\bibinfo  {journal} {Phys. Rev. D}\ }\textbf {\bibinfo
  {volume} {65}},\ \bibinfo {pages} {064035} (\bibinfo {year} {2002})},\
  \bibinfo {note} {[Erratum: Phys.Rev.D 70, 129904 (2004)]},\ \Eprint
  {https://arxiv.org/abs/gr-qc/0108070} {arXiv:gr-qc/0108070} \BibitemShut
  {NoStop}%
\bibitem [{\citenamefont {Le~Tiec}\ and\ \citenamefont
  {Grandcl{\'e}ment}(2018)}]{LeTiec:2017ebm}%
  \BibitemOpen
  \bibfield  {author} {\bibinfo {author} {\bibfnamefont {A.}~\bibnamefont
  {Le~Tiec}}\ and\ \bibinfo {author} {\bibfnamefont {P.}~\bibnamefont
  {Grandcl{\'e}ment}},\ }\bibfield  {title} {\bibinfo {title} {{Horizon Surface
  Gravity in Corotating Black Hole Binaries}},\ }\href
  {https://doi.org/10.1088/1361-6382/aac58c} {\bibfield  {journal} {\bibinfo
  {journal} {Classical Quantum Gravity}\ }\textbf {\bibinfo {volume} {35}},\
  \bibinfo {pages} {144002} (\bibinfo {year} {2018})},\ \Eprint
  {https://arxiv.org/abs/1710.03673} {arXiv:1710.03673 [gr-qc]} \BibitemShut
  {NoStop}%
\bibitem [{\citenamefont {Le~Tiec}\ \emph {et~al.}(2012)\citenamefont
  {Le~Tiec}, \citenamefont {Blanchet},\ and\ \citenamefont
  {Whiting}}]{LeTiec:2011ab}%
  \BibitemOpen
  \bibfield  {author} {\bibinfo {author} {\bibfnamefont {A.}~\bibnamefont
  {Le~Tiec}}, \bibinfo {author} {\bibfnamefont {L.}~\bibnamefont {Blanchet}},\
  and\ \bibinfo {author} {\bibfnamefont {B.~F.}\ \bibnamefont {Whiting}},\
  }\bibfield  {title} {\bibinfo {title} {{The First Law of Binary Black Hole
  Mechanics in General Relativity and Post-Newtonian Theory}},\ }\href
  {https://doi.org/10.1103/PhysRevD.85.064039} {\bibfield  {journal} {\bibinfo
  {journal} {Phys. Rev. D}\ }\textbf {\bibinfo {volume} {85}},\ \bibinfo
  {pages} {064039} (\bibinfo {year} {2012})},\ \Eprint
  {https://arxiv.org/abs/1111.5378} {arXiv:1111.5378 [gr-qc]} \BibitemShut
  {NoStop}%
\bibitem [{\citenamefont {Damour}\ \emph {et~al.}(2012)\citenamefont {Damour},
  \citenamefont {Nagar}, \citenamefont {Pollney},\ and\ \citenamefont
  {Reisswig}}]{Damour:2011fu}%
  \BibitemOpen
  \bibfield  {author} {\bibinfo {author} {\bibfnamefont {T.}~\bibnamefont
  {Damour}}, \bibinfo {author} {\bibfnamefont {A.}~\bibnamefont {Nagar}},
  \bibinfo {author} {\bibfnamefont {D.}~\bibnamefont {Pollney}},\ and\ \bibinfo
  {author} {\bibfnamefont {C.}~\bibnamefont {Reisswig}},\ }\bibfield  {title}
  {\bibinfo {title} {{Energy versus Angular Momentum in Black Hole Binaries}},\
  }\href {https://doi.org/10.1103/PhysRevLett.108.131101} {\bibfield  {journal}
  {\bibinfo  {journal} {Phys. Rev. Lett.}\ }\textbf {\bibinfo {volume} {108}},\
  \bibinfo {pages} {131101} (\bibinfo {year} {2012})},\ \Eprint
  {https://arxiv.org/abs/1110.2938} {arXiv:1110.2938 [gr-qc]} \BibitemShut
  {NoStop}%
\bibitem [{\citenamefont {Nagar}\ \emph {et~al.}(2016)\citenamefont {Nagar},
  \citenamefont {Damour}, \citenamefont {Reisswig},\ and\ \citenamefont
  {Pollney}}]{Nagar:2015xqa}%
  \BibitemOpen
  \bibfield  {author} {\bibinfo {author} {\bibfnamefont {A.}~\bibnamefont
  {Nagar}}, \bibinfo {author} {\bibfnamefont {T.}~\bibnamefont {Damour}},
  \bibinfo {author} {\bibfnamefont {C.}~\bibnamefont {Reisswig}},\ and\
  \bibinfo {author} {\bibfnamefont {D.}~\bibnamefont {Pollney}},\ }\bibfield
  {title} {\bibinfo {title} {{Energetics and phasing of nonprecessing spinning
  coalescing black hole binaries}},\ }\href
  {https://doi.org/10.1103/PhysRevD.93.044046} {\bibfield  {journal} {\bibinfo
  {journal} {Phys. Rev. D}\ }\textbf {\bibinfo {volume} {93}},\ \bibinfo
  {pages} {044046} (\bibinfo {year} {2016})},\ \Eprint
  {https://arxiv.org/abs/1506.08457} {arXiv:1506.08457 [gr-qc]} \BibitemShut
  {NoStop}%
\bibitem [{\citenamefont {Ossokine}\ \emph {et~al.}(2018)\citenamefont
  {Ossokine}, \citenamefont {Dietrich}, \citenamefont {Foley}, \citenamefont
  {Katebi},\ and\ \citenamefont {Lovelace}}]{Ossokine:2017dge}%
  \BibitemOpen
  \bibfield  {author} {\bibinfo {author} {\bibfnamefont {S.}~\bibnamefont
  {Ossokine}}, \bibinfo {author} {\bibfnamefont {T.}~\bibnamefont {Dietrich}},
  \bibinfo {author} {\bibfnamefont {E.}~\bibnamefont {Foley}}, \bibinfo
  {author} {\bibfnamefont {R.}~\bibnamefont {Katebi}},\ and\ \bibinfo {author}
  {\bibfnamefont {G.}~\bibnamefont {Lovelace}},\ }\bibfield  {title} {\bibinfo
  {title} {{Assessing the Energetics of Spinning Binary Black Hole Systems}},\
  }\href {https://doi.org/10.1103/PhysRevD.98.104057} {\bibfield  {journal}
  {\bibinfo  {journal} {Phys. Rev. D}\ }\textbf {\bibinfo {volume} {98}},\
  \bibinfo {pages} {104057} (\bibinfo {year} {2018})},\ \Eprint
  {https://arxiv.org/abs/1712.06533} {arXiv:1712.06533 [gr-qc]} \BibitemShut
  {NoStop}%
\bibitem [{\citenamefont {Boyle}\ \emph {et~al.}(2019)\citenamefont {Boyle}
  \emph {et~al.}}]{Boyle:2019kee}%
  \BibitemOpen
  \bibfield  {author} {\bibinfo {author} {\bibfnamefont {M.}~\bibnamefont
  {Boyle}} \emph {et~al.},\ }\bibfield  {title} {\bibinfo {title} {{The SXS
  Collaboration catalog of binary black hole simulations}},\ }\href
  {https://doi.org/10.1088/1361-6382/ab34e2} {\bibfield  {journal} {\bibinfo
  {journal} {Classical Quantum Gravity}\ }\textbf {\bibinfo {volume} {36}},\
  \bibinfo {pages} {195006} (\bibinfo {year} {2019})},\ \Eprint
  {https://arxiv.org/abs/1904.04831} {arXiv:1904.04831 [gr-qc]} \BibitemShut
  {NoStop}%
\bibitem [{\citenamefont {Rincon-Ramirez}\ \emph {et~al.}(2025)\citenamefont
  {Rincon-Ramirez}, \citenamefont {Johnson-McDaniel}, \citenamefont {Gupta},
  \citenamefont {Prasad}, \citenamefont {Bianchi},\ and\ \citenamefont
  {Sathyaprakash}}]{RinconRamirez:2025maxent}%
  \BibitemOpen
  \bibfield  {author} {\bibinfo {author} {\bibfnamefont {M.}~\bibnamefont
  {Rincon-Ramirez}}, \bibinfo {author} {\bibfnamefont {N.~K.}\ \bibnamefont
  {Johnson-McDaniel}}, \bibinfo {author} {\bibfnamefont {I.}~\bibnamefont
  {Gupta}}, \bibinfo {author} {\bibfnamefont {V.}~\bibnamefont {Prasad}},
  \bibinfo {author} {\bibfnamefont {E.}~\bibnamefont {Bianchi}},\ and\ \bibinfo
  {author} {\bibfnamefont {B.~S.}\ \bibnamefont {Sathyaprakash}},\ }\bibfield
  {title} {\bibinfo {title} {Examining the maximum entropy principle for
  spinning black hole mergers}} (\bibinfo {year} {2025}),\ \bibinfo {note} {to
  be submitted to Phys.\ Rev.\ D}\BibitemShut {NoStop}%
\bibitem [{\citenamefont {Jaramillo}\ \emph {et~al.}(2012)\citenamefont
  {Jaramillo}, \citenamefont {Macedo}, \citenamefont {Moesta},\ and\
  \citenamefont {Rezzolla}}]{Jaramillo:2012rr}%
  \BibitemOpen
  \bibfield  {author} {\bibinfo {author} {\bibfnamefont {J.~L.}\ \bibnamefont
  {Jaramillo}}, \bibinfo {author} {\bibfnamefont {R.~P.}\ \bibnamefont
  {Macedo}}, \bibinfo {author} {\bibfnamefont {P.}~\bibnamefont {Moesta}},\
  and\ \bibinfo {author} {\bibfnamefont {L.}~\bibnamefont {Rezzolla}},\
  }\bibfield  {title} {\bibinfo {title} {{Towards a cross-correlation approach
  to strong-field dynamics in Black Hole spacetimes}},\ }\href
  {https://doi.org/10.1063/1.4734411} {\bibfield  {journal} {\bibinfo
  {journal} {AIP Conf. Proc.}\ }\textbf {\bibinfo {volume} {1458}},\ \bibinfo
  {pages} {158} (\bibinfo {year} {2012})},\ \Eprint
  {https://arxiv.org/abs/1205.3902} {arXiv:1205.3902 [gr-qc]} \BibitemShut
  {NoStop}%
\bibitem [{\citenamefont {Chen}\ \emph {et~al.}(2022)\citenamefont {Chen} \emph
  {et~al.}}]{Chen:2022dxt}%
  \BibitemOpen
  \bibfield  {author} {\bibinfo {author} {\bibfnamefont {Y.}~\bibnamefont
  {Chen}} \emph {et~al.},\ }\bibfield  {title} {\bibinfo {title} {{Multipole
  moments on the common horizon in a binary-black-hole simulation}},\ }\href
  {https://doi.org/10.1103/PhysRevD.106.124045} {\bibfield  {journal} {\bibinfo
   {journal} {Phys. Rev. D}\ }\textbf {\bibinfo {volume} {106}},\ \bibinfo
  {pages} {124045} (\bibinfo {year} {2022})},\ \Eprint
  {https://arxiv.org/abs/2208.02965} {arXiv:2208.02965 [gr-qc]} \BibitemShut
  {NoStop}%
\bibitem [{\citenamefont {Prasad}(2024)}]{Prasad:2024vsz}%
  \BibitemOpen
  \bibfield  {author} {\bibinfo {author} {\bibfnamefont {V.}~\bibnamefont
  {Prasad}},\ }\bibfield  {title} {\bibinfo {title} {{Tidal deformation of
  dynamical horizons in binary black hole mergers and its imprint on
  gravitational radiation}},\ }\href
  {https://doi.org/10.1103/PhysRevD.109.044033} {\bibfield  {journal} {\bibinfo
   {journal} {Phys. Rev. D}\ }\textbf {\bibinfo {volume} {109}},\ \bibinfo
  {pages} {044033} (\bibinfo {year} {2024})}\BibitemShut {NoStop}%
\bibitem [{LAL(2025)}]{LALSuite}%
  \BibitemOpen
  \href {https://doi.org/10.7935/GT1W-FZ16} {\bibinfo {title}
  {{LIGO-Virgo-KAGRA} {A}lgorithm {L}ibrary - {LALS}uite}},\ \bibinfo
  {howpublished} {free software (GPL)} (\bibinfo {year} {2025})\BibitemShut
  {NoStop}%
\bibitem [{\citenamefont {Hunter}(2007)}]{Hunter:2007ouj}%
  \BibitemOpen
  \bibfield  {author} {\bibinfo {author} {\bibfnamefont {J.~D.}\ \bibnamefont
  {Hunter}},\ }\bibfield  {title} {\bibinfo {title} {{Matplotlib: A 2D Graphics
  Environment}},\ }\href {https://doi.org/10.1109/MCSE.2007.55} {\bibfield
  {journal} {\bibinfo  {journal} {Comput. Sci. Eng.}\ }\textbf {\bibinfo
  {volume} {9}},\ \bibinfo {pages} {90} (\bibinfo {year} {2007})}\BibitemShut
  {NoStop}%
\bibitem [{\citenamefont {Harris}\ \emph {et~al.}(2020)\citenamefont {Harris}
  \emph {et~al.}}]{Harris:2020xlr}%
  \BibitemOpen
  \bibfield  {author} {\bibinfo {author} {\bibfnamefont {C.~R.}\ \bibnamefont
  {Harris}} \emph {et~al.},\ }\bibfield  {title} {\bibinfo {title} {{Array
  programming with NumPy}},\ }\href {https://doi.org/10.1038/s41586-020-2649-2}
  {\bibfield  {journal} {\bibinfo  {journal} {Nature (London)}\ }\textbf
  {\bibinfo {volume} {585}},\ \bibinfo {pages} {357} (\bibinfo {year}
  {2020})},\ \Eprint {https://arxiv.org/abs/2006.10256} {arXiv:2006.10256
  [cs.MS]} \BibitemShut {NoStop}%
\bibitem [{\citenamefont {Boyle}\ \emph {et~al.}(2025)\citenamefont {Boyle},
  \citenamefont {Mitman}, \citenamefont {Scheel},\ and\ \citenamefont
  {Stein}}]{sxs_package}%
  \BibitemOpen
  \bibfield  {author} {\bibinfo {author} {\bibfnamefont {M.}~\bibnamefont
  {Boyle}}, \bibinfo {author} {\bibfnamefont {K.}~\bibnamefont {Mitman}},
  \bibinfo {author} {\bibfnamefont {M.~A.}\ \bibnamefont {Scheel}},\ and\
  \bibinfo {author} {\bibfnamefont {L.~C.}\ \bibnamefont {Stein}},\ }\href
  {https://doi.org/10.5281/zenodo.4034006} {\bibinfo {title} {The sxs package}}
  (\bibinfo {year} {2025})\BibitemShut {NoStop}%
\bibitem [{\citenamefont {Damour}\ \emph {et~al.}(1998)\citenamefont {Damour},
  \citenamefont {Iyer},\ and\ \citenamefont {Sathyaprakash}}]{Damour:1997ub}%
  \BibitemOpen
  \bibfield  {author} {\bibinfo {author} {\bibfnamefont {T.}~\bibnamefont
  {Damour}}, \bibinfo {author} {\bibfnamefont {B.~R.}\ \bibnamefont {Iyer}},\
  and\ \bibinfo {author} {\bibfnamefont {B.~S.}\ \bibnamefont
  {Sathyaprakash}},\ }\bibfield  {title} {\bibinfo {title} {{Improved filters
  for gravitational waves from inspiralling compact binaries}},\ }\href
  {https://doi.org/10.1103/PhysRevD.57.885} {\bibfield  {journal} {\bibinfo
  {journal} {Phys. Rev. D}\ }\textbf {\bibinfo {volume} {57}},\ \bibinfo
  {pages} {885} (\bibinfo {year} {1998})},\ \Eprint
  {https://arxiv.org/abs/gr-qc/9708034} {arXiv:gr-qc/9708034} \BibitemShut
  {NoStop}%
\end{thebibliography}%

\end{document}